\documentclass[aps,pre,twocolumn,groupedaddress,showpacs]{revtex4}

\usepackage[T1]{fontenc}
\usepackage[utf8]{inputenc}

\usepackage{amsmath}
\usepackage{amsfonts}
\usepackage{amssymb}
\usepackage{hyperref}

\usepackage{graphicx}
\graphicspath{{./}{./figures/}{./figures-eps/}{./figures-pdf/}}

\begin{document}
\title{A large-scale study of the World Wide Web: \\
network correlation functions with scale-invariant boundaries}

\author{G.A. Ludue\~{n}a}
\affiliation{Institute for Theoretical Physics, Goethe University Frankfurt, Germany} 
\author{H. Meixner}
\affiliation{Institute for Theoretical Physics, Goethe University Frankfurt, Germany} 
\author{Gregor Kaczor}
\affiliation{Institute for Theoretical Physics, Goethe University Frankfurt, Germany} 
\author{Claudius Gros}
\affiliation{Institute for Theoretical Physics, Goethe University Frankfurt, Germany}


\date{\today}


\begin{abstract}
We performed a large-scale crawl of the World Wide Web, 
covering $6.9$ Million domains, including all high-traffic 
sites of the Internet. We present a study of the correlations 
found between quantities measuring the structural relevance of each node 
in the network (the in- and out-degree, the local clustering coefficient, 
the first-neighbor in-degree and the Alexa rank). We find that some of these 
properties show strong correlation effects and that the dependencies 
occurring out of these correlations follow power laws not only for
the averages, but also for the boundaries of the respective 
density distributions. 
In addition, these scale-free limits do not follow the same exponents as the 
corresponding averages. In our study we retain the directionality of the
hyperlinks and develop a statistical estimate for the clustering coefficient
of directed graphs.

We include in our study the correlations between the in-degree 
and the Alexa traffic rank, a popular index for the traffic volume, 
finding non-trivial power-law correlations. We find that sites 
with more/less than about $10^3$ links from different domains 
have remarkably different statistical properties, for all correlation 
functions studied, indicating towards an underlying hierarchical
structure of the World Wide Web.
\end{abstract}

\maketitle

\section{Introduction}

The emergence of the World Wide Web (WWW) belongs arguably to the 
most relevant events of the present time. The interest in this 
system and in networks in general permeated through all the society, 
including physics. This led, at the turn of the century, to a large 
amount of studies of what with the time came to be known as 
``network science''. Most studies of the WWW were performed,
however, in the early 2000s 
\cite{webstructure2000,barabasi-albert,dogorostev-mendes}
and large-scale studies of the WWW are rather hard 
to find nowadays, despite the immense growth of the Internet 
in the last 10 years.

A remarkable finding of the first generation studies of the WWW
is the emergence of scale-free degree distributions, which can be
explained potentially from the view of preferential attachment, 
although the exponents obtained are not universal \cite{dogorostev-mendes}. 
Generally, one can assume that the growth process of a complex network
will be influenced by inter-node correlations and that these dependencies
will be reflected in the resulting network topology. However, such 
correlations are not easy to detect and characterize, and
have not been studied in depth. It is expected that a simple rule as
preferential attachment cannot completely reproduce the structures found in
real-world networks, and therefore more complicated models have been developed
to replicate the behavior \cite{connectivityOfRandomNets,pref-link-struc,
correls-internet,makse-selfsim,varying-vertex-fitness}.

Correlations between different properties are generally 
used as a proxy to study the internal structure of the 
network. For instance, Vespagnani studied correlations 
between the in-degree of a node and that of a first neighbor 
of said node \cite{correls-internet}, showing a scale free 
property (recently modeled by Takagi \cite{takagi2012}), 
Barabasi and Albert studied the local clustering coefficient 
as a function of the in-degree \cite{hierarchy-barabasi},
in order to obtain information regarding the hierarchical 
structuring of the network. However, real-world data about 
said correlations is not abundant.

In the present work we study the complete dominant core of the WWW
by crawling $6.9$ Million domains, including all domains with the 
largest traffic (all domains with an Alexa rank of one Million 
or less are included). Collapsing the data, by neglecting
link multiplicities, we study the network of inter-domain 
hyperlinks (not webpages), containing about half a Billion 
directed edges. We find non-trivial correlations between 
in- and out-degree, between the in-degree and the local 
clustering coefficient and between the degrees of 
neighboring sites. In addition to evaluate averaged quantities,
we study the full density plots, finding novel scaling features 
for the boundaries of several correlation functions. We present,
in addition, a formula for the clustering coefficient of random 
directed graph characterized by given arbitrary in- and out-degree 
sequences. Finally we present an analysis of the correlations between the
number of in-links and the Alexa rank of a domain.

\begin{figure}[t]
\centering
\noindent
\includegraphics[height=0.11\textwidth]{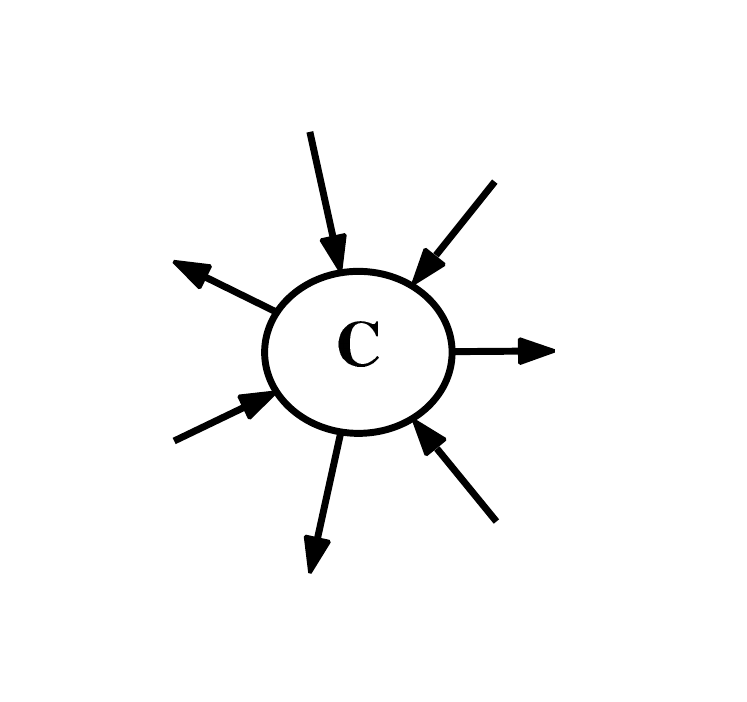}
\hspace{4ex}
\includegraphics[height=0.11\textwidth]{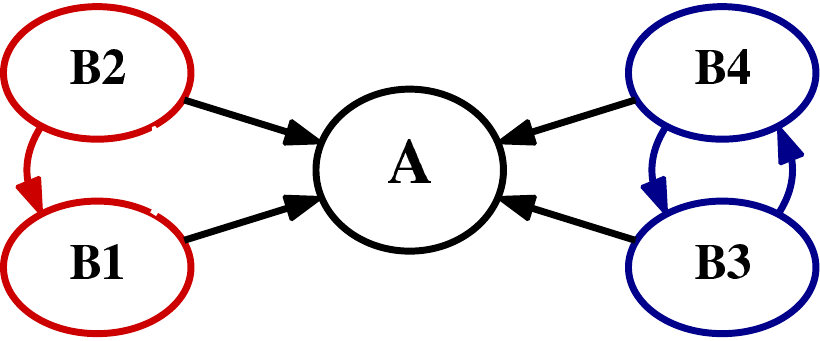}
\caption{Left: A node with in-degree $k_{in}=4$ and
out-degree $k_{out}=3$. Right: Two types of in-degree clusters,
with the edges always directed towards the central site (A).} 
\label{fig:graph-illustrations}
\end{figure}

\section{Theory}

For directed graphs we have to distinguish between the
distribution $p_{in}(k)$ and $p_{out}(l)$ of the
in- and the out-degrees $k$ and $l$, respectively. There are, 
in addition, two kinds of nearest neighbors, in-neighbors
and out-neighbors. Site B1 in Fig.~\ref{fig:graph-illustrations}
is a nearest in-neighbor of site A and a nearest out-neighbor 
of site B2. Alternatively one could call B1 an ancestor of 
A and a descendent of B2 \cite{gros-book}. For bi-directional
links, as between B3 and B4 in Fig.~\ref{fig:graph-illustrations},
in-neighbors are also out-neighbors. The total number of in-links 
equals the total number of out-links, the in- and out-degree 
coordination numbers 
\begin{equation}
z \ =\ \sum_k k\,p_{in}(k) \ =\ \sum_l l\,p_{out}(l)
\label{def_z}
\end{equation}
are hence identical.

\subsection{Clustering coefficient model for directed graphs}

In order to calculate the relevance of correlations between 
in- and out- degree in the structure of the network, we have 
developed a statistical model of the clustering coefficient 
for given distributions of in- and out- degree which are
uncorrelated.

We define with
\begin{equation}
q_{out}(l) \ =\  \frac{(l+1)\,p_{out}(l+1)}{N_q}
\label{def_q-out}
\end{equation}
the excess distribution \cite{newman2003social} of 
outgoing links of a nearest in-neighbor. The normalization
constant $N_q=\sum_l l\,p_{out}(l)$ is
just the coordination number $z$, see (\ref{def_z}).
Equivalently we define via
\begin{equation}
\bar q_{in}(k)  \ =\  
\frac{1}{N_{\bar q}}\sum_l p_{in,out}(k,l)\,l
\label{def_bar-q-in}
\end{equation}
the degree distribution (not excess) of incoming links of a 
nearest in-neighboring site. Here $p_{in,out}(k,l)$
is the probability that a site has $l$ out-links 
and $k$ in-links (joint distribution function),
with the usual relations 
\begin{eqnarray}
	\sum_k p_{in,out}(k,l) &=& p_{out}(l), \\
\sum_l p_{in,out}(k,l) &=& p_{in}(k)
\end{eqnarray}
for the marginal distribution functions. 
The normalization constant $N_{\bar q}$ in 
(\ref{def_bar-q-in}) is given by the coordination 
number $z$,
$$
N_{\bar q} \ =\  \sum_k\sum_l p_{in,out}(k,l)\,l 
    \ =\ \sum_l l\,p_{out}(l) \ \equiv\  z \; .
$$
For the clustering coefficient $\hat C$ (the `hat' 
symbol stands here for the clustering coefficient 
of a directed graph) we now consider two in-neighbors, 
having respectively, with probabilities $\bar q_{in}(k)$
and $q_{out}(l)$, $k$ in-links and $l$ excess out-links 
(as stubs).

We now assume that the distributions $q_{in}(k)$
and $\bar{q}_{out}(l)$ of the two neighbors are independent 
of each other. The probability, for a graph with $N$ nodes,
 that a given pair of in- and out-stubs are connected is 
then $1/(Nz)$, where $Nz$ is the total number of in- or 
out-stubs, and hence
\begin{eqnarray*}
\hat C & =& \frac{1}{Nz}\sum_{k,l} \bar q_{in}(k)\,k\,l\,q_{out}(l)\\
  & =& \frac{1}{Nz^3}
\left(\sum_{k}k\,p_{in,out}(k,l) l \right) \!
\left(\sum_{l}p_{out}(l+1) l (l+1) \right).
\end{eqnarray*}
Transforming now into a sum $\sum_s$ over sites, 
every site $s$ being characterized by an in-degree $k_s$ 
and out-degree $l_s$, one obtains
\begin{equation}
\hat C \ =\ \frac{1}{Nz^3}
\left(\frac{1}{N}\sum_{s}k_sl_s \right) 
\left(\frac{1}{N}\sum_{s}(l_s-1)l_s \right)~,
\label{eq:hat-C-sites}
\end{equation}
which coincides with the usual expression \cite{gros-book}
for non-directed graphs (apart from a factor 
$l_s$ instead of $l_s-1$ in the first factor),
by taking $k_s=l_s$ for $s=1,\dots,N$. A fully-connected 
network results in $\hat C=1$ under this formula.

We note that the expression (\ref{eq:hat-C-sites})
for $\hat C$ may actually violate the sum rule
$\hat C\le 1$, due to the neglect of inter-site
degree correlations, when applied to a real-world graph. 
As an example consider a network composed out of a single 
star, like the site C in Fig.~\ref{fig:graph-illustrations}, 
but with bi-directional edges. For a un-directed (and
loopless) star the degree sequence is
$$
k_1=l_1=N-1, 
\qquad
k_i=l_i=1,
\qquad i=2,\dots, N~,
$$
with an intensive coordination number
$z=2(N-1)/N\approx 2$. The statistical formula 
(\ref{eq:hat-C-sites}) for the clustering coefficient 
would, one the other hand, diverge
$$
\approx\frac{1}{N2^3}\left(\frac{1}{N}\Big( (N-1)^2 + (N-1)\Big)
              \right)^2 \ \sim\ \frac{N}{2^3}
$$
in the thermodynamic limit $N\to\infty$. A substantial
deviation of $\hat C$  from the true clustering coefficient
is hence a measure for the strength of inter-site
degree correlations, the expression (\ref{eq:hat-C-sites})
being valid for graphs with vanishing inter-node
correlations.

\begin{figure}[t]
\centering
\includegraphics[width=0.45\textwidth]{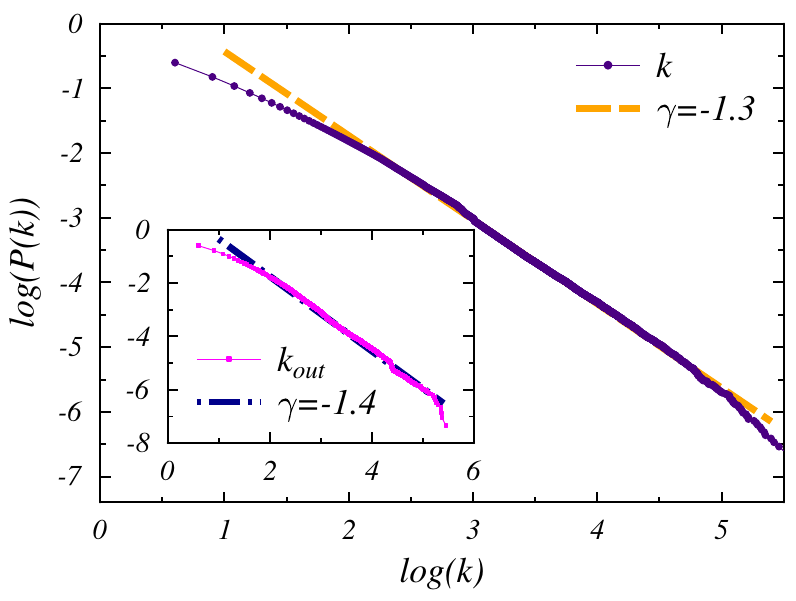}
\caption{Complementary cumulative distributions 
$P(k)=\int_x^\infty p(k')dk'$ for the in-degree (main panel)
and the out-degree (inset), log-log plot. The dashed lines,
corresponding to power-law distributions, have slopes
$\gamma=-1.3$ and $\gamma=-1.4$, respectively for the in- and
the out-degree.} 
\label{fig:in-out-degree-distribution}
\end{figure}

\section{Results}

Using the crawlers of the former file search engine
FindFiles.net \cite{gros2012neuropsychological} 
we crawled, mostly in 2011, $6.9$ Million domains (of
type {\tt http://www.domain.com}) with a total of
$64$ Million subdomains (of type {\tt http://subdomain.domain.com}). 
These $6.9$ Million domains have 223 Million hyperlinks in between 
them, linking in addition to 50 Million other sites.  For the 
network analysis we neglected these 50 Million external sites,
as we did not crawl them separately. The network of 223 Million 
inter-domain directed links has an average degree of 32 and
$0.7$ Million of the $6.9$ Million domains are isolated in 
the sense that they have no in-links, they cannot be reached 
from the core of the World Wide Web. A further one Million 
sites have just a single hyperlink directed to them.

The crawling strategy started from the set of the about 32 Million 
subdomains referred-to in Wikipedia and DMOZ (all languages), with 
further systematic additional extensions. We included, in particular,
the one Million domains with the largest traffic volume, in terms
of the Alexa rank. This data set, which we denote with FF-2011, 
hence corresponds essentially to the complete relevant part of the 
World Wide Web, in terms of traffic volume.

\subsection{In- and out-degree distributions}

The degree distribution of hyperlinks have been 
observed to follow a power law $\sim k^{\gamma} $, with 
an exponent close to the limiting case $\gamma\to -2$
(when the mean degree would diverge in the thermodynamic
limit) \cite{barabasi-albert,albert-diameter-web-nature,
barabasi-revmod,dogorostev-mendes}. In 
Fig.~\ref{fig:in-out-degree-distribution} we present the
complementary cumulative distribution functions 
\cite{markovic2013} for both the in-degree and the 
out-degree.

Over a range of about 2.5-3 orders of magnitude,
the data can be approximated quite nicely by power law
distributions, with exponents $\gamma_{in}=-2.3$ and 
$\gamma_{out}=-2.4$ respectively for the in- and the 
out-degree. These results confirm earlier studies 
\cite{barabasi2000scale,broder2000graph,gros2012neuropsychological} 
finding consistently $|\gamma_{in}|<|\gamma_{out}|$. The
absolute magnitude of the values reported for the
scaling exponents vary slightly from study to study, 
either because of the evolution of the Internet with 
time passing, or due to the size of the respective databases.

\begin{figure}[t]
\centering
\includegraphics[width=0.45\textwidth]{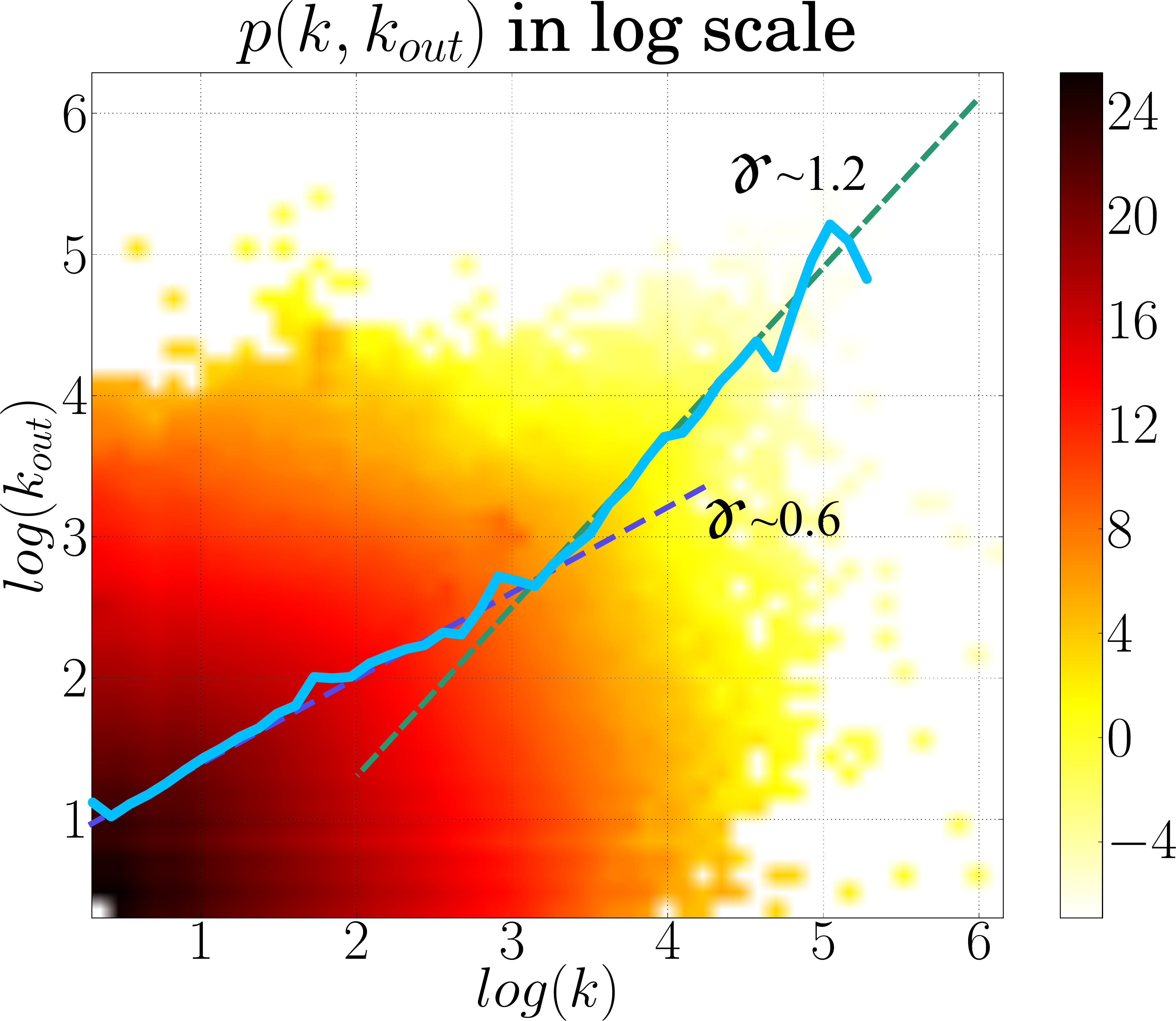}
\caption{Density distribution $p_{in,out}(k,k_{out})$
of domains with in-degree $k$ and out-degree $k_{out}$. 
The density is shown in log scale, as well as both axes. 
The solid line represents the average $\langle k_{out}\rangle(k)$.
The probability density is normalized, 
$\int \int p_{in,out}(k,k_{out}) dk dk_{out} = 1$.}
\label{fig:kVsKout}
\end{figure}

\subsection{Correlations between in- and out-degree}

In Fig.~\ref{fig:kVsKout}, the density distribution of nodes having
an in-degree $k$ and an out-degree $k_{out}$, is presented, together
with the average out-degree $\langle k_{out} \rangle(k)$, for
sites having an in-degree $k$. In- and out-degree do not seem 
to be particularly correlated, on a first sight. However, the
average out-degree shows two regimes with approximated 
power-law scaling, for $k<10^3$ and $k>10^3$, with
exponents $\gamma=-0.6$ and $\gamma=-1.2$ respectively.
In the case that the joint distribution 
$p_{in,out}(k,k_{out})$ would factorize,
$p_{in,out}(k,k_{out})\to p_{in}(k)p_{out}(k_{out})$,
the mean out-degree
\begin{equation}
\langle k_{out} \rangle(k) 
\ =\ \int p(k,l) \,l\,dl
\ \to\ p_{in}(k)\, z~,
\label{eq:in_out}
\end{equation}
would functionally follow the in-degree distribution $p_{in}(k)$,
where $z\approx32$ is the average (in- and out-) 
degree of our Internet data. However, as shown
in Fig.~\ref{fig:in-out-degree-distribution},
the marginal in-degree distribution $p_{in}(k)$,
falls approximately like $k^{-2.3}$, viz 
substantially faster than (\ref{eq:in_out}) would
imply. In- and out-degree are hence non-trivially
correlated. We will discuss the nature of the
respective correlations in more detail further below
when discussing the distribution of local
clustering coefficients.

\subsection{Mean clustering coefficient}

The local clustering coefficient $C_i$ is given by
the number of linked nearest neighbors of site $i$,
relative to the total number of possible links between
the neighbors. For directed graphs there are
in- and out-neighbors and various possible 3-site loops,
as illustrated in Fig.~\ref{fig:graph-illustrations},
also known as network motifs \cite{milo2002network,alon2007network}.
Here we examine the in-clustering coefficient.
For a given site $i$ the in-clustering coefficient
$C_i$ is given by the average number of links in between
the in-neighbors of site $i$. In Fig.~\ref{fig:graph-illustrations},
the sites (B1,A,B2) form an in-loop of site A, contributing
to $C_A$, while the sites (B4,A,B3) contribute two in-loops.
We focus on the in-clustering coefficient since
the number of in-links is a measure for the
importance of a site, contributing to its traffic
volume.

We find, for the FF-2011 network data, a mean
clustering coefficient $\bar C=\sum_i C_i/N$ 
of $\bar C = 0.18$. This is, for two reasons,
a surprising high value. Firstly the connection
probability $p$ is very low, being just
$p=4.6\times 10^{-6}$. Secondly a quite large number
of sites, 0.27\%, has a vanishing local clustering 
$C_i=0$, and only a small fraction, 0.3\%, of domains, 
mostly with small degrees, have a maximal local 
clustering coefficient of unity.

We can assess the impact of correlations on the
formation of local loops by considering identical
degree sequences for the in- and out- degree, as
extracted from the FF-2011 network data, but 
considering various types of correlations between 
the in- and out- degree of each node.

\begin{itemize}
\item
Applying Eq.~(\ref{eq:hat-C-sites}) to the actual network, 
the obtained value amounts to $\hat C_{model}=1.5$. 
This value is over the maximum $C=1$, indicating towards
very strong correlations between the in- and the out-degree
distributions, compare the discussion below
Eq.~(\ref{eq:hat-C-sites}).

\item For a network having the same degree distributions 
$p_{in}(k)$ and $p_{out}(l)$ for the in- and the out-degree
as the actual network, but without correlations between these 
degrees, viz assuming a joint probability distribution
$p_{in,out}(k,l)\to p_{in}(k)p_{out}(l)$,
the clustering coefficient obtained by Eq.~(\ref{eq:hat-C-sites}) 
would amount to $\hat C_{decorr}=2.4\times 10 ^{-3}$. 

\item A network where the in- and out- degrees are
      anticorrelated (nodes with largest in-degree are mapped to
      the smallest out-degree), would amount to an even lower 
      $\hat C_{anticorr}=3.2\times 10 ^{-4}$. 

\item For a network with a maximally correlated distribution of 
      in- and out- degree (nodes with the largest in-degree being
      mapped to the largest out-degree), would result again in a
      higher-than-unity clustering coefficient 
      $\hat C_{maxcorr}=3.2$, when using  
      Eq.~(\ref{eq:hat-C-sites}).
\end{itemize}

We hence conclude that the in- and out-degree are quite
strongly correlated positively for the World Wide Web.

\begin{figure}[t]
\centering
\includegraphics[width=0.45\textwidth]{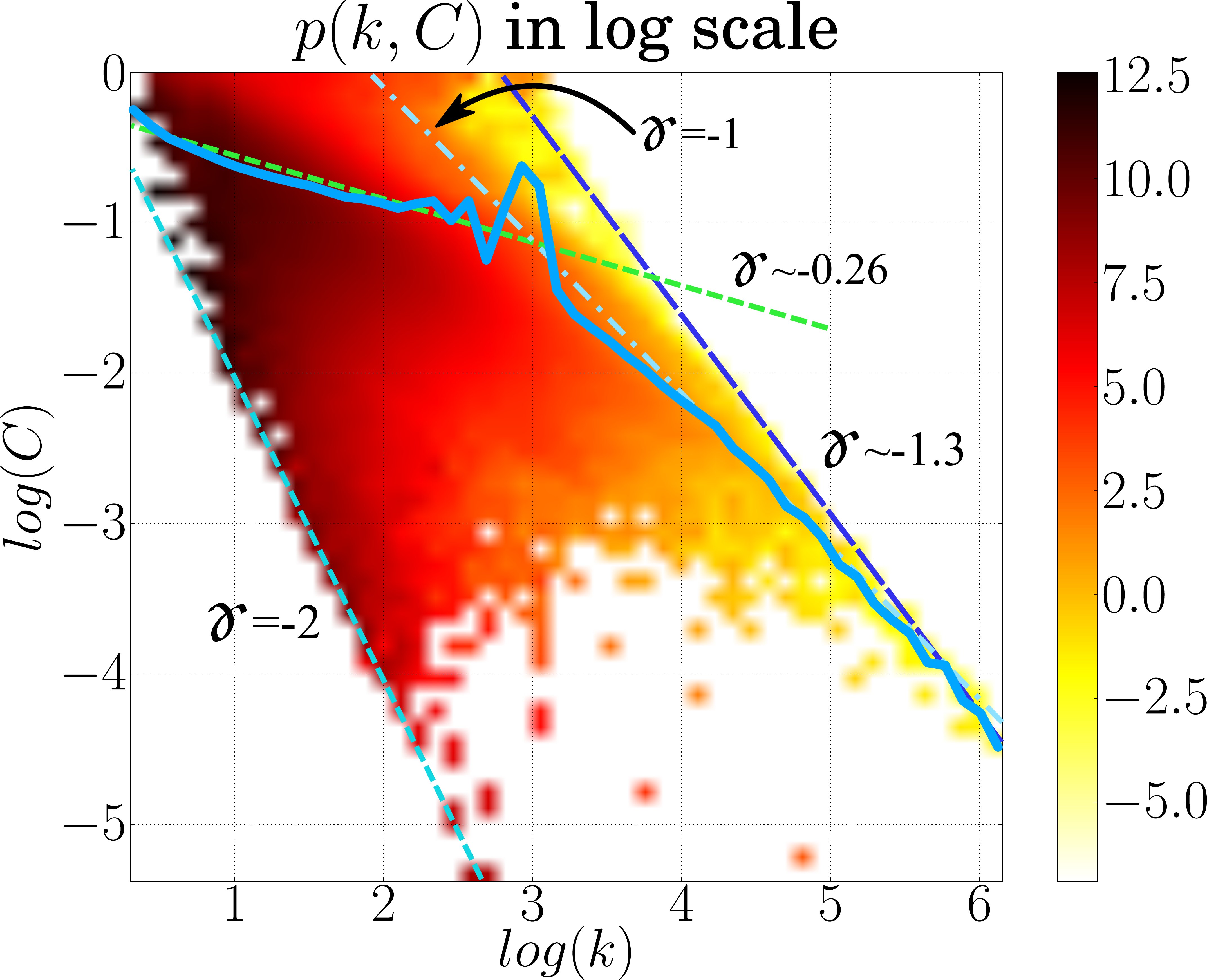}
\caption{Probability density $P(k,C)$ of pairs of
in-degree $k$ and local clustering coefficient $C$.
The density, and both axes are given in log scales.
The solid line represents the average value 
$\langle C\rangle(k)$, as a function of in-degree $k$. The 
probability density is normalized, 
$\int \int P(k,C) dk dC = 1$.}
\label{fig:kVsC}
\end{figure}

\subsection{Distribution of local clustering coefficients}

In Fig.~\ref{fig:kVsC}, the density of $(k,C)$ pairs is shown in a 
log-log scale, where $k$ is the in-degree and $C$ the local
clustering coefficient. The density distribution has 
upper and lower cutoffs scaling approximatively like
$\sim k^{\gamma}$, with $\gamma_{max}\approx -1.3$ and 
$\gamma_{min}=-2$. The lower limit has a simple explanation. 
The lowest non-zero local clustering coefficient is realized
when just a single loop exist out of the $k(k-1)$ 
possible triangles,
\begin{equation}
C_{min}\ =\ \frac{n_l}{k(k-1)} 
	\ \propto\  k^{-2} ~,
\label{eq:C_min}
\end{equation}
when setting the number of loops $n_l$ to one. The exponent of 
the upper limit, $\gamma_{max}=-1.3$, implies, compare (\ref{eq:C_min}),
that the number of local loops scales like $\sim k^{0.7}$. We
have presently no explanation for this scaling behavior.

The average value of $\langle C \rangle(k)$,  as a 
function of in-degree $k$, follows mostly a power law for 
small $k < 10^3$, with an exponent $\gamma=-0.26$. For 
larger $k> 10^4$ the exponent changes toward $\gamma= -1$ 
for the mean local clustering coefficient. This last 
exponent is in agreement with previous observations found in
\cite{hierarchy-barabasi}, and are a fingerprint for a 
hierarchical network structure. The change in behavior at 
the point $k=10^3$ is also observable in the correlation
between the in-degree and the degree of nearest neighbors,
as we will show in the next sections.

There is a group of nodes with very high clustering 
coefficients $C\simeq 1$ around the $k\sim 10^3$ region 
(close to where the upper limit with the $\gamma_{max}=-1$
slope intersects the abscissa), which somewhat falls of the
line. After analyzing some of the domains involved in this 
region, we conclude that this group of nodes does not
represent the intrinsic network structure of the WWW,
belonging most probably to \emph{link farms}. These nodes
are however responsible for the jumps in $\langle C \rangle(k)$ 
at $k\sim 10^3$.

\begin{figure}[t]
\centering
\includegraphics[width=0.45\textwidth]{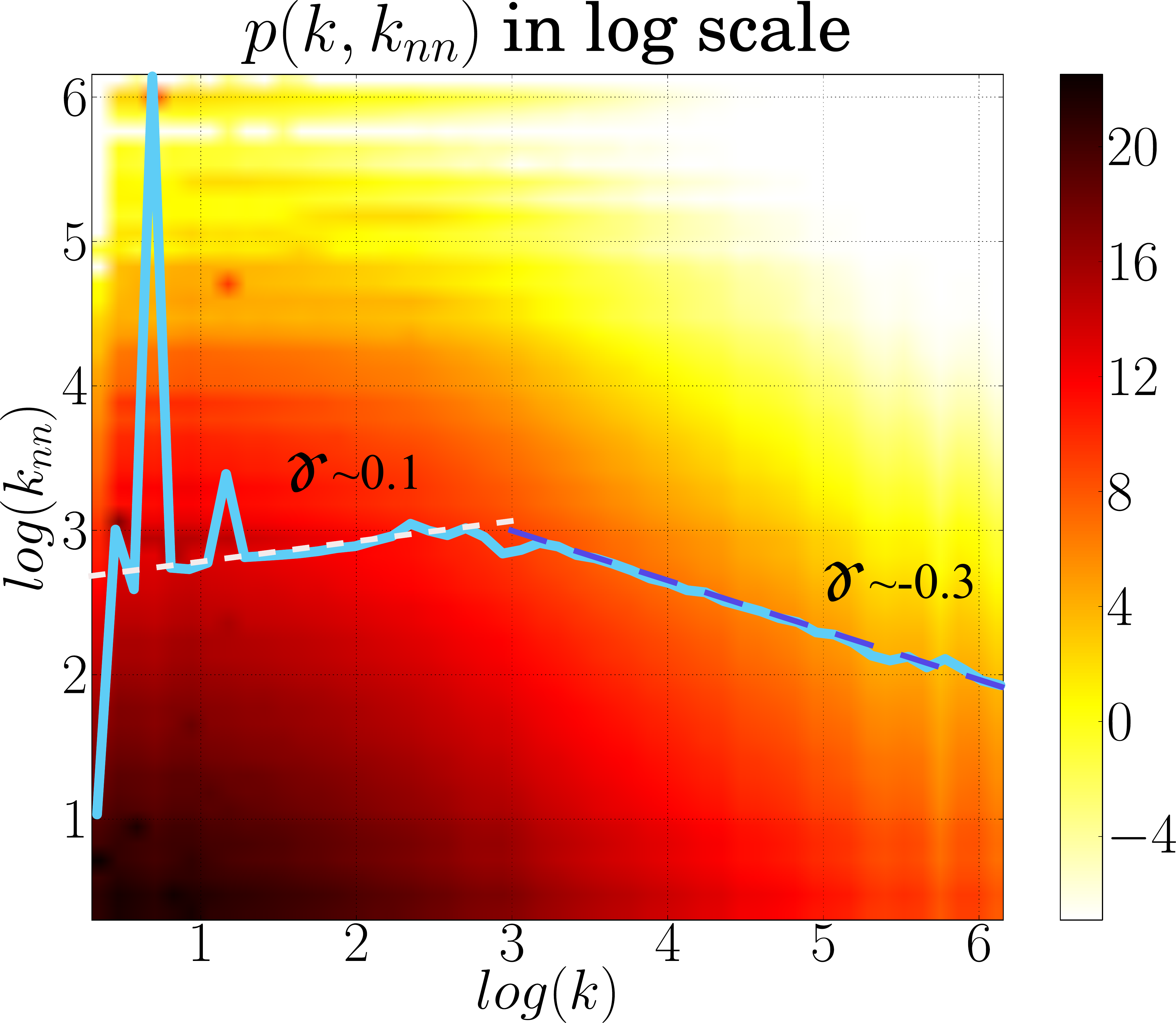}
\caption{The probability density distribution 
$P(k,k_{nn})$ of the in-degree $k$ and the in-degree
$k_{nn}$ of nearest neighbor sites. Both the density
distribution and the axis are given in log scale.
The solid line corresponds to the average value 
$\langle k_{nn}\rangle(k)$. The probability distribution
is normalized, such that 
$\int \int P(k,k_{nn}) dk dk_{nn} = 1$.}
\label{fig:kvsknn}
\end{figure}

\subsection{Nearest-neighbor degree correlations}

In Fig.~\ref{fig:kvsknn}, the density of pairs $(k,k_{nn})$ 
is shown, where $k_{nn}$ is the in-degree of a first neighbor,
and $k$ the in-degree.
The dots with higher densities for low $k$ values are 
relatively large groups of linked domains which share 
exactly the same pairs of in-degree and first-neighbor 
in-degree. The domains do not seem to be particularly related
although we do not discard the possibility that they may 
belong to \emph{link farms}, as they clearly stand out of 
the general behavior of the density distribution.

When analyzing the average $\langle k_{nn} \rangle(k)$ 
as a function of the in-degree $k$, we observe a very 
weak increase for small $k$ until $k\approx 10^3$. We
can fit this increase fairly good with a power law of 
exponent $\gamma\approx 0.1$. This behavior would be 
in agreement with the one observed in the canonical 
Barabasi-Albert model
\cite{barabasi-albert,albert-diameter-web-nature}, 
though it differs with with a 1998 WWW network study \cite{albert-diameter-web-nature}.

In the range from $k\approx 10^3$ to $k\approx 10^6$ we observe 
a change in the behavior of the average $\langle k_{nn} \rangle(k)$, 
as it starts decaying with increasing $k$. This decay follows a 
power-law as well, with an exponent of about $\gamma \approx -0.3$. 
This decay is closer to the results found in
\cite{albert-diameter-web-nature} for a subset of the 1998 
Internet data and the fitness model developed therein 
(which decays with $\gamma=-0.5$). However, the decay is 
observed in our results for much higher degree $k$ than in 
\cite{albert-diameter-web-nature}, which has data limited 
to $k\le10^3$. We speculate that this difference is due to 
the size of the network studied, although it might be possible 
that the evolution of the WWW in the last 10 years is responsible 
for the structural change.

\begin{figure}[t]
\centering
\includegraphics[width=0.45\textwidth]{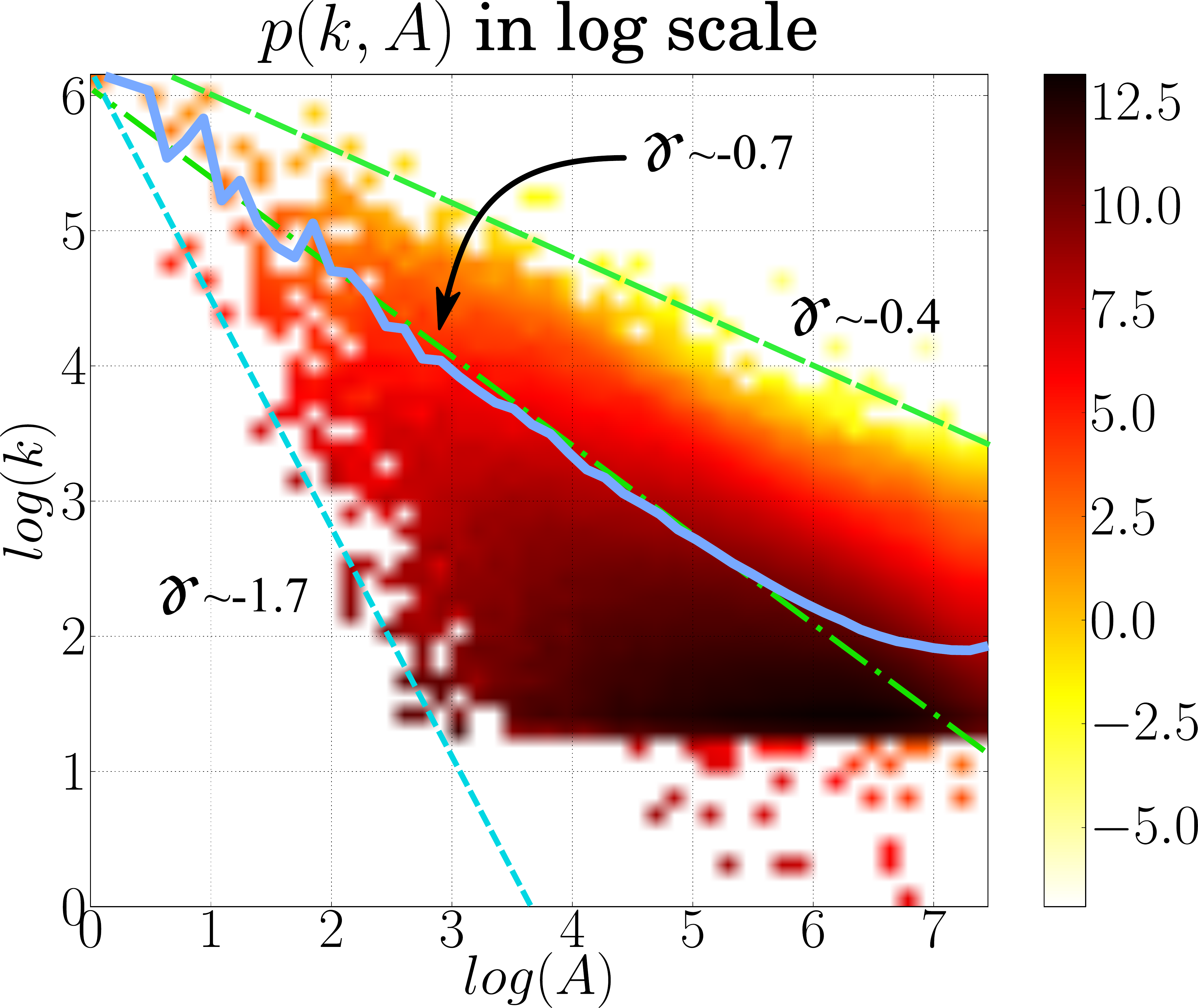}
\caption{Density of pairs $(A,k)$, where $A$ is the Alexa index and
$k$ the corresponding in-degree of the domain. The density is given
in log scale, as well as both axes. The solid line shows the 
average $\langle k\rangle(A)$, as a function of $A$. 
The probability is normalized, such that $\int \int P(k,A)
dk dA = 1$.}
\label{fig:alexaVsK}
\end{figure}

\subsection{Correlations between in-degree and Alexa index}

We have analyzed the correlations of the Alexa rank 
\cite{alexa} with respect to the in-degree $k$. The 
Alexa rank is arguably one of the most popular measures
of the traffic received by an Internet site, and so, 
its relevance. The ranking is proprietary, so the general 
public does not have access to the specifics of its 
calculation, although according to the official information,
it is derived from the traffic observed, with data partly
retrieved from users who installed the Alexa add-on to their web
browser \cite{alexa-howto}. In this ranking, the site with 
the most traffic has rank $A=1$, the following largest rank 
$A=2$ and so on. The rank does not provide any information 
about the precise amount of traffic, such that a larger $A$
index does not give any indication of how much less traffic 
does that site receives, but rather only that it receives 
less traffic than the sites with smaller $A$.

\begin{figure*}[t]
\centering
\includegraphics[width=.48\textwidth]{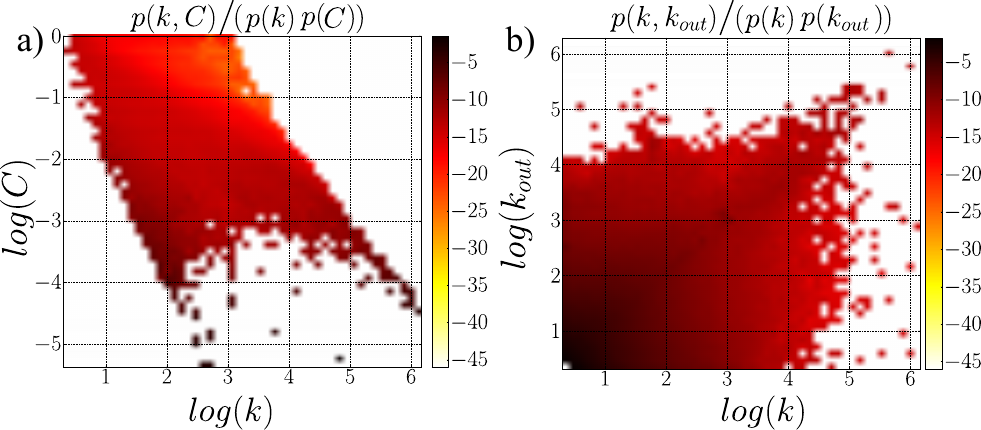}
\includegraphics[width=.48\textwidth]{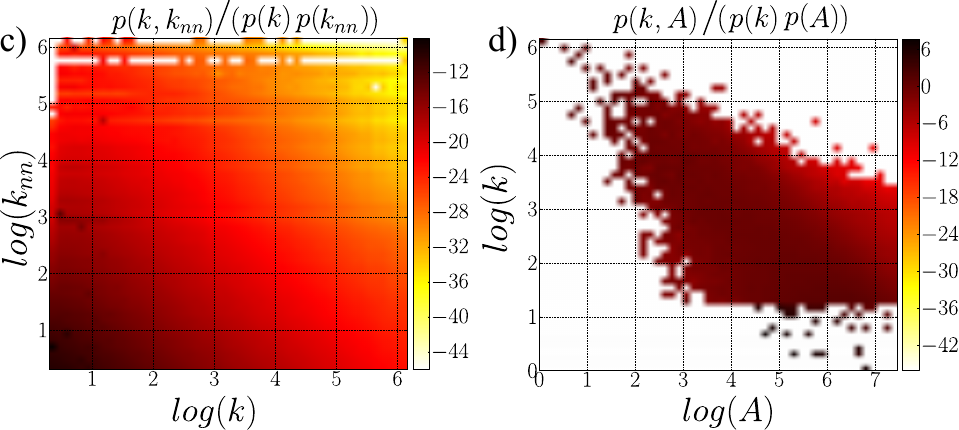}
\caption{Probability distributions normalized by their marginal distributions: a)
$\frac{p(k,C)}{p(k)p(C)}$, b) $\frac{p(k,k_{out})}{p(k)p(k_{out})}$, c)
$\frac{p(k,k_{nn})}{p(k)p(k_{nn})}$, d) $\frac{p(k,A)}{p(k)p(A)}$, all
densities in log scale.} \label{fig:correl}
\end{figure*}

In Fig.~\ref{fig:alexaVsK}, we present the density of 
domains as a function of its in-degree $k$ and Alexa rank 
$A$. We only analyze the Alexa rank for sites having an
in-degree $k>20$, with a few exceptions, due to constraints
in retrieving the Alexa rank data. We observe a 
distribution limited from above and below by two power 
laws with exponents $\gamma_{upper}=-0.4$ and 
$\gamma_{lower}\approx -1.7$. The lower limit is however
less pronounced, due to the lack of samples.

The solid line in Fig.~\ref{fig:alexaVsK} shows the 
average in-degree $\langle k\rangle(A)$, for sites having
and Alexa rank $A$. We observe a very marked power 
law decay
\begin{equation}
\langle k \rangle \ \sim \frac{1}{A^{0.7}},
\qquad\quad
\langle A \rangle \ \sim \frac{1}{k^{1.8}}~.
\label{eq:scaling_k-A}
\end{equation}
The exponents are not the inverse of each other, since
$\langle A\rangle$ and $\langle k\rangle$ are distinct
averages. There is a saturation at $k\sim10^6$ for the scaling regime, 
presumable due to our constraint $k>20$ for the Alexa index.
We find this scaling particularly interesting, since 
the Alexa rank is not derived directly from the 
topology of the network but rather from the traffic
generated by users. For site administrators the
relatively weak decay (\ref{eq:scaling_k-A}) implies
that the traffic generated by in-coming hyperlinks
can be a relevant contribution to the overall traffic
volume.

\subsection{Normalization Studies}

We have performed a visualization study of the
data discussed hitherto by considering relative
joint distribution functions, which are obtained
by dividing a given joint probability distribution
by the product of the respective marginal distributions,
$$
\frac{p(x,y)}{p(x)p(y)},
\quad p(x) =\int p(x,y)dy,
\quad p(y) =\int p(x,y)dx~.
$$
In the absence of correlations, viz when
$p(x,y)\to p(x)p(y)$, the respective density 
plots would be homogeneous and flat.

In Fig.~\ref{fig:correl}a, we show the relative density 
$p(k,C)/(p(k)p(C))$ for pairs of in-degree $k$ and local clustering
coefficient $C$. The distribution is quite homogeneous and 
the upper and lower limits of the distribution are 
exalted in comparison with the plot of the bar 
joint distribution presented in Fig.~\ref{fig:kVsC}.

In Fig.~\ref{fig:correl}b, we present the relative joint density 
for the correlation between in- and out-degree.
The distribution is considerably more homogeneous
than the respective bare probability density shown in
Fig.~\ref{fig:kVsKout}. However, a substantial enhancement
remains for small in- and out- degrees. 

From the shape of the joint nearest-neighbor degree
distribution presented in In Fig.~\ref{fig:kvsknn},
it would be tempting to think that its shape is mostly 
determined by the marginal distributions, i.e.\ that
$p(k,k_{nn}) \approx p(k)p(k_{nn})$, and that therefore $k$ 
and $k_{nn}$ would be essentially decorrelated. This is, however, 
not the case, as we can see in Fig.~\ref{fig:correl}c. In this 
plot the correlations are seen more clearly in terms of
the relative joint distribution We can observe that the 
resulting relative distribution still shows a stronger 
correlation when both in-degree and the in-degree of the 
neighbor are small.

In Fig.~\ref{fig:correl}d, we present the relative density 
distribution of for sites having an Alexa rank $A$ and an 
in-degree $k$. As for the case of the clustering coefficient 
$C$, the distribution maintains the very marked upper and 
lower limits, being otherwise essentially only 
slightly more uniform than the orginal data shown in
Fig.~\ref{fig:alexaVsK}.

\section{Links from $10^3$ distinct domains}

The present study shows that many properties of the WWW are characterized
by non-trivial correlations. We observe that the joint probability 
distributions, for several of the properties tested, follow power-law 
scaling for the respective averages. Additionally, the distributions have, in many
instances, density distributions which are limited by power laws. The power
law limiting functions do not share exponents neither with that of the marginal
distributions, nor with the respective average value of the property studied. 
Interestingly one also observes power-law scaling for a seemingly unrelated 
quantity, the Alexa traffic rank, which decays as a function of
in-degree and also the average in-degree decays moderately weakly as
a function of the Alexa rank.

We found that the statistical properties of the World Wide Web differ
remarkably for domains receiving more/fewer than about $10^3$ hyperlinks
from different domains. The change in behavior is observed for the
correlations between in- and out-degree, between in-degree and local
clustering coefficient and between in-degree and the in-degree of neighbors.
This observation points towards an underlying hierarchical structure of the WWW,
with the ``elite'' of the Internet domains, receiving links from more than 
one thousand different domains, being made-up by about $20\cdot10^3$ sites.

\section{Acknowledgements}
We would like to acknowledge the support of the German Science Foundation
(DFG).


\begin{thebibliography}{22}
\expandafter\ifx\csname natexlab\endcsname\relax\def\natexlab#1{#1}\fi
\expandafter\ifx\csname bibnamefont\endcsname\relax
  \def\bibnamefont#1{#1}\fi
\expandafter\ifx\csname bibfnamefont\endcsname\relax
  \def\bibfnamefont#1{#1}\fi
\expandafter\ifx\csname citenamefont\endcsname\relax
  \def\citenamefont#1{#1}\fi
\expandafter\ifx\csname url\endcsname\relax
  \def\url#1{\texttt{#1}}\fi
\expandafter\ifx\csname urlprefix\endcsname\relax\def\urlprefix{URL }\fi
\providecommand{\bibinfo}[2]{#2}
\providecommand{\eprint}[2][]{\url{#2}}

\bibitem[{\citenamefont{Broder et~al.}(2000{\natexlab{a}})\citenamefont{Broder,
  Kumar, Maghoul, Raghavan, Rajagopalan, Stata, Tomkins, and
  Wiener}}]{webstructure2000}
\bibinfo{author}{\bibfnamefont{A.}~\bibnamefont{Broder}},
  \bibinfo{author}{\bibfnamefont{R.}~\bibnamefont{Kumar}},
  \bibinfo{author}{\bibfnamefont{F.}~\bibnamefont{Maghoul}},
  \bibinfo{author}{\bibfnamefont{P.}~\bibnamefont{Raghavan}},
  \bibinfo{author}{\bibfnamefont{S.}~\bibnamefont{Rajagopalan}},
  \bibinfo{author}{\bibfnamefont{R.}~\bibnamefont{Stata}},
  \bibinfo{author}{\bibfnamefont{A.}~\bibnamefont{Tomkins}}, \bibnamefont{and}
  \bibinfo{author}{\bibfnamefont{J.}~\bibnamefont{Wiener}},
  \bibinfo{journal}{Computer Networks} \textbf{\bibinfo{volume}{33}},
  \bibinfo{pages}{309} (\bibinfo{year}{2000}{\natexlab{a}}).

\bibitem[{\citenamefont{Barabási and Albert}(1999)}]{barabasi-albert}
\bibinfo{author}{\bibfnamefont{A.-L.} \bibnamefont{Barabási}}
  \bibnamefont{and} \bibinfo{author}{\bibfnamefont{R.}~\bibnamefont{Albert}},
  \bibinfo{journal}{Science} \textbf{\bibinfo{volume}{286}},
  \bibinfo{pages}{509} (\bibinfo{year}{1999}).

\bibitem[{\citenamefont{Dogorostev and Mendes}(2003)}]{dogorostev-mendes}
\bibinfo{author}{\bibfnamefont{S.~N.} \bibnamefont{Dogorostev}}
  \bibnamefont{and} \bibinfo{author}{\bibfnamefont{J.~F.~F.}
  \bibnamefont{Mendes}}, \emph{\bibinfo{title}{Evolution of Networks}}
  (\bibinfo{publisher}{Oxford University Press}, \bibinfo{year}{2003}).

\bibitem[{\citenamefont{Krapivsky et~al.}(2000)\citenamefont{Krapivsky, Redner,
  and Leyvraz}}]{connectivityOfRandomNets}
\bibinfo{author}{\bibfnamefont{P.~L.} \bibnamefont{Krapivsky}},
  \bibinfo{author}{\bibfnamefont{S.}~\bibnamefont{Redner}}, \bibnamefont{and}
  \bibinfo{author}{\bibfnamefont{F.}~\bibnamefont{Leyvraz}},
  \bibinfo{journal}{Phys. Rev. Lett.} \textbf{\bibinfo{volume}{85}},
  \bibinfo{pages}{4629} (\bibinfo{year}{2000}).

\bibitem[{\citenamefont{Dorogovtsev et~al.}(2000)\citenamefont{Dorogovtsev,
  Mendes, and Samukhin}}]{pref-link-struc}
\bibinfo{author}{\bibfnamefont{S.~N.} \bibnamefont{Dorogovtsev}},
  \bibinfo{author}{\bibfnamefont{J.~F.~F.} \bibnamefont{Mendes}},
  \bibnamefont{and} \bibinfo{author}{\bibfnamefont{A.~N.}
  \bibnamefont{Samukhin}}, \bibinfo{journal}{Phys. Rev. Lett.}
  \textbf{\bibinfo{volume}{85}}, \bibinfo{pages}{4633} (\bibinfo{year}{2000}).

\bibitem[{\citenamefont{Pastor-Satorras
  et~al.}(2001)\citenamefont{Pastor-Satorras, V\'azquez, and
  Vespignani}}]{correls-internet}
\bibinfo{author}{\bibfnamefont{R.}~\bibnamefont{Pastor-Satorras}},
  \bibinfo{author}{\bibfnamefont{A.}~\bibnamefont{V\'azquez}},
  \bibnamefont{and}
  \bibinfo{author}{\bibfnamefont{A.}~\bibnamefont{Vespignani}},
  \bibinfo{journal}{Phys. Rev. Lett.} \textbf{\bibinfo{volume}{87}},
  \bibinfo{pages}{258701} (\bibinfo{year}{2001}).

\bibitem[{\citenamefont{Song et~al.}(2005)\citenamefont{Song, Havlin, and
  Makse}}]{makse-selfsim}
\bibinfo{author}{\bibfnamefont{C.}~\bibnamefont{Song}},
  \bibinfo{author}{\bibfnamefont{S.}~\bibnamefont{Havlin}}, \bibnamefont{and}
  \bibinfo{author}{\bibfnamefont{H.~A.} \bibnamefont{Makse}},
  \bibinfo{journal}{Nature} \textbf{\bibinfo{volume}{433}}, \bibinfo{pages}{392
  } (\bibinfo{year}{2005}).

\bibitem[{\citenamefont{Caldarelli et~al.}(2002)\citenamefont{Caldarelli,
  Capocci, De~Los~Rios, and Mu\~noz}}]{varying-vertex-fitness}
\bibinfo{author}{\bibfnamefont{G.}~\bibnamefont{Caldarelli}},
  \bibinfo{author}{\bibfnamefont{A.}~\bibnamefont{Capocci}},
  \bibinfo{author}{\bibfnamefont{P.}~\bibnamefont{De~Los~Rios}},
  \bibnamefont{and} \bibinfo{author}{\bibfnamefont{M.~A.}
  \bibnamefont{Mu\~noz}}, \bibinfo{journal}{Phys. Rev. Lett.}
  \textbf{\bibinfo{volume}{89}}, \bibinfo{pages}{258702}
  (\bibinfo{year}{2002}).

\bibitem[{\citenamefont{Takagi}(2012)}]{takagi2012}
\bibinfo{author}{\bibfnamefont{K.}~\bibnamefont{Takagi}},
  \bibinfo{journal}{World Journal of Mechanics} \textbf{\bibinfo{volume}{2}},
  \bibinfo{pages}{171} (\bibinfo{year}{2012}).

\bibitem[{\citenamefont{Ravasz and Barab\'asi}(2003)}]{hierarchy-barabasi}
\bibinfo{author}{\bibfnamefont{E.}~\bibnamefont{Ravasz}} \bibnamefont{and}
  \bibinfo{author}{\bibfnamefont{A.-L.} \bibnamefont{Barab\'asi}},
  \bibinfo{journal}{Phys. Rev. E} \textbf{\bibinfo{volume}{67}},
  \bibinfo{pages}{026112} (\bibinfo{year}{2003}).

\bibitem[{\citenamefont{Gros}(2008)}]{gros-book}
\bibinfo{author}{\bibfnamefont{C.}~\bibnamefont{Gros}},
  \emph{\bibinfo{title}{Complex and Adaptive Dynamical Systems, a Primer.}}
  (\bibinfo{publisher}{Springer, New York}, \bibinfo{year}{2008}).

\bibitem[{\citenamefont{Newman and Park}(2003)}]{newman2003social}
\bibinfo{author}{\bibfnamefont{M.}~\bibnamefont{Newman}} \bibnamefont{and}
  \bibinfo{author}{\bibfnamefont{J.}~\bibnamefont{Park}},
  \bibinfo{journal}{Physical Review E} \textbf{\bibinfo{volume}{68}},
  \bibinfo{pages}{036122} (\bibinfo{year}{2003}).

\bibitem[{\citenamefont{Gros et~al.}(2012)\citenamefont{Gros, Kaczor, and
  Markovi{\'c}}}]{gros2012neuropsychological}
\bibinfo{author}{\bibfnamefont{C.}~\bibnamefont{Gros}},
  \bibinfo{author}{\bibfnamefont{G.}~\bibnamefont{Kaczor}}, \bibnamefont{and}
  \bibinfo{author}{\bibfnamefont{D.}~\bibnamefont{Markovi{\'c}}},
  \bibinfo{journal}{The European Physical Journal B-Condensed Matter and
  Complex Systems} \textbf{\bibinfo{volume}{85}}, \bibinfo{pages}{1}
  (\bibinfo{year}{2012}).

\bibitem[{\citenamefont{Albert et~al.}(1999)\citenamefont{Albert, Jeong, and
  Barabasi}}]{albert-diameter-web-nature}
\bibinfo{author}{\bibfnamefont{R.}~\bibnamefont{Albert}},
  \bibinfo{author}{\bibfnamefont{H.}~\bibnamefont{Jeong}}, \bibnamefont{and}
  \bibinfo{author}{\bibfnamefont{A.-L.} \bibnamefont{Barabasi}},
  \bibinfo{journal}{Nature} \textbf{\bibinfo{volume}{401}},
  \bibinfo{pages}{130} (\bibinfo{year}{1999}).

\bibitem[{\citenamefont{Albert and Barab\'asi}(2002)}]{barabasi-revmod}
\bibinfo{author}{\bibfnamefont{R.}~\bibnamefont{Albert}} \bibnamefont{and}
  \bibinfo{author}{\bibfnamefont{A.-L.} \bibnamefont{Barab\'asi}},
  \bibinfo{journal}{Rev. Mod. Phys.} \textbf{\bibinfo{volume}{74}},
  \bibinfo{pages}{47} (\bibinfo{year}{2002}).

\bibitem[{\citenamefont{Markovic and Gros}(2013)}]{markovic2013}
\bibinfo{author}{\bibfnamefont{D.}~\bibnamefont{Markovic}} \bibnamefont{and}
  \bibinfo{author}{\bibfnamefont{C.}~\bibnamefont{Gros}}, \bibinfo{journal}{(to
  be published)}  (\bibinfo{year}{2013}).

\bibitem[{\citenamefont{Barab{\'a}si et~al.}(2000)\citenamefont{Barab{\'a}si,
  Albert, and Jeong}}]{barabasi2000scale}
\bibinfo{author}{\bibfnamefont{A.}~\bibnamefont{Barab{\'a}si}},
  \bibinfo{author}{\bibfnamefont{R.}~\bibnamefont{Albert}}, \bibnamefont{and}
  \bibinfo{author}{\bibfnamefont{H.}~\bibnamefont{Jeong}},
  \bibinfo{journal}{Physica A: Statistical Mechanics and its Applications}
  \textbf{\bibinfo{volume}{281}}, \bibinfo{pages}{69} (\bibinfo{year}{2000}).

\bibitem[{\citenamefont{Broder et~al.}(2000{\natexlab{b}})\citenamefont{Broder,
  Kumar, Maghoul, Raghavan, Rajagopalan, Stata, Tomkins, and
  Wiener}}]{broder2000graph}
\bibinfo{author}{\bibfnamefont{A.}~\bibnamefont{Broder}},
  \bibinfo{author}{\bibfnamefont{R.}~\bibnamefont{Kumar}},
  \bibinfo{author}{\bibfnamefont{F.}~\bibnamefont{Maghoul}},
  \bibinfo{author}{\bibfnamefont{P.}~\bibnamefont{Raghavan}},
  \bibinfo{author}{\bibfnamefont{S.}~\bibnamefont{Rajagopalan}},
  \bibinfo{author}{\bibfnamefont{R.}~\bibnamefont{Stata}},
  \bibinfo{author}{\bibfnamefont{A.}~\bibnamefont{Tomkins}}, \bibnamefont{and}
  \bibinfo{author}{\bibfnamefont{J.}~\bibnamefont{Wiener}},
  \bibinfo{journal}{Computer networks} \textbf{\bibinfo{volume}{33}},
  \bibinfo{pages}{309} (\bibinfo{year}{2000}{\natexlab{b}}).

\bibitem[{\citenamefont{Milo et~al.}(2002)\citenamefont{Milo, Shen-Orr,
  Itzkovitz, Kashtan, Chklovskii, and Alon}}]{milo2002network}
\bibinfo{author}{\bibfnamefont{R.}~\bibnamefont{Milo}},
  \bibinfo{author}{\bibfnamefont{S.}~\bibnamefont{Shen-Orr}},
  \bibinfo{author}{\bibfnamefont{S.}~\bibnamefont{Itzkovitz}},
  \bibinfo{author}{\bibfnamefont{N.}~\bibnamefont{Kashtan}},
  \bibinfo{author}{\bibfnamefont{D.}~\bibnamefont{Chklovskii}},
  \bibnamefont{and} \bibinfo{author}{\bibfnamefont{U.}~\bibnamefont{Alon}},
  \bibinfo{journal}{Science Signalling} \textbf{\bibinfo{volume}{298}},
  \bibinfo{pages}{824} (\bibinfo{year}{2002}).

\bibitem[{\citenamefont{Alon}(2007)}]{alon2007network}
\bibinfo{author}{\bibfnamefont{U.}~\bibnamefont{Alon}},
  \bibinfo{journal}{Nature Reviews Genetics} \textbf{\bibinfo{volume}{8}},
  \bibinfo{pages}{450} (\bibinfo{year}{2007}).

\bibitem[{ale({\natexlab{a}})}]{alexa}
\urlprefix\url{www.alexa.com}.

\bibitem[{ale({\natexlab{b}})}]{alexa-howto}
\urlprefix\url{http://www.alexa.com/company/technology}.

\end{thebibliography}



\end{document}